\documentstyle[preprint,aps,epsfig]{revtex}

\def\be{\begin{equation}}
\def\ee{\end{equation}}
\def\ep{\epsilon}
\def\t{\tilde}

\begin{document}
\draft
\title{Nonlinear ac response of anisotropic composites}
\author{J. P. Huang, Jones T. K. Wan, C. K. Lo and K. W. Yu}
\address{Department of Physics, The Chinese University of Hong Kong, \\
         Shatin, New Territories, Hong Kong}
\maketitle

\begin{abstract}
When a suspension consisting of dielectric particles having nonlinear
characteristics is subjected to a sinusoidal (ac) field, 
the electrical response will in general consist of ac fields at 
frequencies of the higher-order harmonics. 
These ac responses will also be anisotropic.
In this work, a self-consistent formalism has been employed to
compute the induced dipole moment for suspensions in which the
suspended particles have nonlinear characteristics, 
in an attempt to investigate the anisotropy in the ac response.
The results showed that the harmonics of the induced dipole moment
and the local electric field are both increased as the anisotropy
increases for the longitudinal field case, while the harmonics are 
decreased as the anisotropy increases for the transverse field case.
These results are qualitatively understood with the spectral representation.
Thus, by measuring the ac responses both parallel and perpendicular to 
the uniaxial anisotropic axis of the field-induced structures, 
it is possible to perform a real-time monitoring of the field-induced 
aggregation process.
\end{abstract}
\vskip 5mm 

\section{Introduction}
When a suspension of highly polarizable dielectric particles is subjected
to an intense electric field, the induced dipole moments cause the 
particles to form chains along the field direction, resulting in complex
anisotropic structures. These chain structures alter the apparent viscosity 
of the suspension by several orders of magnitude, and are the basis of the
electrorheological (ER) effect. 
Under the action of the ER effect, the arrangement of particles, or the 
shape of the lattice changes and the local electric field deviates
from the Lorentz cavity field \cite{Lo}. 
As a result, these structures have anisotropic physical properties, 
such as in the effective conductivity and permittivity \cite{Martin-2}, 
and in the optical nonlinerity enhancement \cite{Yuen}.
The rapid field-induced aggregation and the large anistropy in 
their properties render this field-induced-structured material potentially 
important for technological applications \cite{ER}.

Moreover, the applied electric field used in most ER experiments is 
quite high, and important data on nonlinear ER effects induced 
by a strong electric field have been reported recently \cite{Kling98}. 
A convenient method of probing the nonlinear characteristics of the 
field-induced structures is to measure the harmonics of the nonlinear 
polarization under the application of a sinusoidal (ac) electric field. 
Thus by measuring the ac responses both parallel and perpendicular to 
the uniaxial anisotropic axis of the structures, it is possible to
perform a real-time monitoring of the field-induced aggregation process.

Recently, the effect of nonlinear characteristics on the interparticle 
force has been analyzed in an ER suspension of nonlinear particles 
\cite{nler-1} and further extended to a nonlinear host medium 
\cite{nler-2}. 
The nonlinear characteristics are due to the field-dependence of the
dielectric constants of the materials used in ER fluids, which have a 
constitutive relation ${\bf D}=\ep {\bf E} + \chi {\bf E}^3$. 
When a nonlinear composite with nonlinear dielectric particles embedded
in a host medium, or with a nonlinear host medium is subjected to
a sinusoidal field, the electrical response in the composite will
in general be a superposition of many sinusoidal functions
\cite{Bergman95,Hui}. It is natural to investigate the effects of a
nonlinear characteristics on the ac response in an ER fluid which can be 
regarded as a nonlinear composite medium \cite{Gu00}. 
The strength of the nonlinear polarization should be
reflected in the magnitude of the harmonics.

In this work, we will develop a perturbation expansion \cite{Gu92} 
method and a self-consistent theory \cite{Yu96} to calculate the 
ac response of a nonlinear compsite. 
The paper is organized as follows. In the next section, we calculate the
dipole moment of the polarized spherical particles in an anisotropic
composite and extract the harmonic response. 
In section III, we perform a series expansion of the local field inside 
the particles from the self-consistent solution, in an attempt to obtain
analytic expressions of the higher harmonics of the dipole moment.
Numerical results are performed in section IV to validate the
analytic results. Discussions on the results will be given.

\section{Nonlinear polarization and its higher harmonics}
We first examine the effect of a nonlinear characteristics on the
induced dipole moment. We concentrate on the case where the
suspended particles have a nonlinear dielectric constant, while
the host medium has a linear dielectric constant $\ep_2$. The
nonlinear characteristics gives rise to a field-dependent
dielectric coefficient \cite{Yu93}. In which case, the electric
displacement-electric field relation inside the spheres is given by:
 \be
{\bf D}_1=\ep_1{\bf E}_1+\chi\langle E_1^2\rangle{\bf E}_1
  =\t{\ep}_1{\bf E}_1,
\label{nl-coefficient}
 \ee 
where $\ep_1$ and $\chi$ are the linear coefficient and the
nonlinear coefficient of the suspended particles respectively.

This constitutes an approximation: the local field inside the
particles is assumed to be uniform and the assumption is called
the decoupling approximation \cite{Yu96}. It has been shown that
such an approximation yields a lower bound for the accurate result
for the local field \cite{Yu96}. We further assumed that both
$\ep$ and $\chi$ are independent of frequency, which is a
valid assumption for low-frequency processes in ER fluids. 
As a result, the induced dipole moment under an applied field 
${\bf E}=E(t)\hat{\bf z}$ is given by:
 \be
\t{p}=\t{\ep}_e a^3 \t{b} E(t),
 \ee 
where $\t{b}$ is the field-dependent dipolar factor and is given by:
 \be
\t{b}={\t{\ep}_1-\ep_2 \over
  \t{\ep}_1+2\ep_2} ={\ep_1+\chi\langle
  E_1^2 \rangle-\ep_2 \over \ep_1+\chi\langle E_1^2
  \rangle+2\ep_2}.
 \ee 

In order to obtain the effective dielectric constant $\t{\ep}_e$, we 
invoke the Maxwell-Garnett approximation (MGA) for anisotropic 
composites \cite{Lo}.
For the longitudinal field case when the ac field is applied along
the uniaxial anisotropic axis, the MGA takes on the form
 \be 
\frac{\t{\ep}_e-\ep_2}{\beta_{L}\t{\ep}_e+(3-\beta_{L})\ep_2}=
  f\frac{\t{\ep}_1-\ep_2}{\t{\ep}_1+2\ep_2},
 \ee 
whereas for an transverse field case when the ac field is applied 
perpendicular to the uniaxial anisotropic axis, the MGA reads
  \be
\frac{\t{\ep}_e-\ep_2}{\beta_{T}\t{\ep}_e+(3-\beta_{T})\ep_2}=
  f\frac{\t{\ep}_1-\ep_2}{\t{\ep}_1+2\ep_2} ,
 \ee 
where $\beta_{L}$ and $\beta_{T}$ denote the local field factors parallel
and perpendicular to the uniaxial anisotropic axis respectively,
and $f$ is the volume fraction of particles.
These factors are defined as the ratio of the local field in the 
particles to the Lorentz cavity field \cite{Lo}.
For isotropic composites, $\beta_L=\beta_T=1$, while both $\beta_L$ and 
$\beta_T$ will deviate from unity for an anisotropic distribution of
particles in composites.
The $\beta$ factors have been evaluated in a tetragonal lattice of
dipole moments \cite{Lo} and similar quantities were calculated in 
various field-induced-structured composites \cite{Martin-3}; 
they satisfy the sum rule:
$$
\beta_{L}+2\beta_{T}=3.
$$
In what follows, we denote $\beta_{L}$ as $\beta$ for notation convenience.

On the other hand, the effective nonlinear dielectric constant can be 
given by \cite{Yu93} 
 \be
\t{\ep}_e=\frac{1}{E^2(t)V}\int_V\ep({\bf r})|{\bf E}({\bf r},t)|^2
  {\rm d}V=\frac{f\t{\ep}_1}{E^2(t)}\langle E_1^2\rangle
  +\frac{(1-f)\ep_2}{E^2(t)}\langle E_2^2\rangle ,
 \ee 
where $V$ the volume of the composite and $f$ is the volume fraction. 
Accordingly, the local electric field inside the spheres can be expressed 
in terms of the derivative of $\t{\ep}_e$ with respect to $\t{\ep}_1$:
 \be
\langle E_1^2\rangle=\frac{1}{f}E^2(t)
  \frac{\partial\t{\ep}_e}{\partial\t{\ep}_1}.
 \ee 

If we apply a sinusoidal electric field, i.e. $E(t)=E_0\sin(\omega t)$, 
the induced dipole moment $\t{p}$ will depend on time sinusoidally, too.
By virtue of the inversion symmetry, $\t{p}$ is a superposition of 
odd-order harmonics such that 
 \be
\t{p}=p_{\omega}\sin \omega t+p_{3\omega}\sin 3\omega t+ \cdots.
 \ee 
Also, the local electric field contains similar harmonics
 \be
\sqrt{\langle E_1^2\rangle}=E_{\omega}\sin \omega t+E_{3\omega}\sin 
3\omega t+ \cdots.
 \ee 
These harmonic coefficients can be extracted from the time dependence
of the solution of $\t{p}$ and $E_1(t)$.

\section{Analytic Solutions}
In what follows, we will apply two methods to extract the harmonics of 
the induced dipole moment and the eletric field: the perturbation
expansion method (PEM) \cite{Gu92,Gu00} and the self-consistent (SC) 
theory \cite{Yu96}. 
The self-consistent theory can deal with the case of strong nonlinearity, 
while the perturbation expansion method is applicable to weak 
nonlinearity only, i.e. $\chi\langle E_1^2\rangle\ll 1$, 
limited by the convergence of the series expansion.
In the case of weakly nonlinear response, the SC theory should agree with 
PEM.
 
  \subsection{Perturbation expansion method}

We expand $\t{p}$ and $\sqrt{\chi\langle E_1^2\rangle}$ into a Taylor 
expansion
\begin{eqnarray}
\t{p}&=&a^3E(t)\sum_{n=0}^{\infty}a_n(\chi\langle E_1^2\rangle)^n ,\\
  \sqrt{\chi\langle E_1^2\rangle}&=&\frac{\sqrt{\chi E^2(t)}}
  {f^{1/2}}\sum_{m=0}^{\infty}d_m(\chi\langle E_1^2\rangle)^m ,
\end{eqnarray}
where
$$
a_n = \frac{1}{n!}\frac{\partial^n}{\partial\t{\ep}_1^n}
(\t{\ep}_e\t{b})|_{\t{\ep}_1=\ep_1},\ \ \
d_m = \frac{1}{m!}\frac{\partial^m}{\partial\t{\ep}_1^m}
\sqrt{\frac{\partial \ep_e}{\partial\t{\ep}_1}}|_{\t{\ep}_1=\ep_1}.
$$

For weak nonlinearity, we can rewrite Eqs.(10) and 
(11), keeping the lowest orders of $\chi E^2(t)$ and $\chi\langle 
E_1^2\rangle$:
\begin{eqnarray}
\sqrt{\chi\langle E_1^2\rangle}&=&\frac{\sqrt{\chi E^2(t)}}{f^{1/2}}
 (d_0+d_1\chi \langle E_1^2\rangle) , \\
\t{p}&=&a^3a_0E(t)+\frac{1}{f}a^3a_1d_0^2\chi E^3(t)\equiv 
h_1E(t)+h_3\chi E^3(t). 
\end{eqnarray}

In the case of a sinusoidal field, we can expand $E^3(t)$ in terms of 
the first and the third harmonics.
The comparison with Eq.(8) yields the harmonics of the induced dipole 
moment: 
$$
p_{\omega} = h_1 E_0+\frac{3}{4}h_3\chi E_0^3, \ \ \
p_{3\omega} = -\frac{1}{4}h_3 \chi E_0^3.
$$

Similarly, we find the harmonics of the local electric field
$$
\chi^{1/2}E_{\omega} = j_1\chi^{1/2}E_0+\frac{3}{4}j_3(\chi^{1/2}E_0)^3,
\ \ \
\chi^{1/2}E_{3\omega} = -\frac{1}{4}j_3(\chi^{1/2}E_0)^3 ,
$$
where
$$
j_1 = d_0/f^{1/2},\ \ \
j_3 = d_0^2d_1/f^{3/2}.
$$

In the above analysis, we have used the identity 
$\sin^3\omega t={3\over 4}\sin \omega t- {1\over 4} \sin 3\omega t$ 
to obtain the first and the third harmonics. 
Similar analysis can be used to extract the higher-order harmonics, 
by retaining more terms in the series expansion.

  \subsection{Self-consistent theory}
  
Eq.(7) is actually a self-consistent equation since its right-hand side 
depends on the local field itself.
It must be solved self-consistently, and the local electric field reads:
 \be
\sqrt{\chi \langle E_1^2\rangle}=\frac{-2(3)^{1/3} R+2^{1/3} 
  (9Q\sqrt{T}+\sqrt{3}\sqrt{4R^3+27Q^2 T})^{2/3}}
  {6^{2/3}\sqrt{T}(9Q\sqrt{T}+\sqrt{3}\sqrt{4 R^3+27Q^2T})^{1/3}}
 \ee 
where
$$
Q = 3\ep_2 \chi^{1/2}E(t),\ \ \
R = \ep_1 T+\ep_2 (2+f\beta),\ \ \
T = 1-f \beta.
$$

For a sinusoidal applied electric field, we determine numerically 
the harmonics of the induced dipole moment and the local electric field,
which is a subject for the next section.
 
\section{Numerical Results}
In this section, we perform numerical calculations to investigate
the effects of a nonlinear characteristics on the harmonics of the
induced dipole moment and the local electric field. 
As shown in Section III, the harmonics of the induced dipole moment and 
the local electric field are both related to the field-induced anisotropy 
parameter $\beta$.
In order to investigate the effect of field-induced anisotropy on the
harmonics, we perform the numerical calculations.
Without loss of generality, we let $f=0.09$, $a=1$, $\ep_1=10$, and 
$\ep_2=1$ for model calculations. We let $p_0=\ep_e a^3 b E_0$
to normalize the induced dipole moment.
  
In Fig.1, we plot the normalized harmonics versus $\beta$ by using the SC 
theory for the longitudinal electric field. It is evident that, 
for increasing $\beta$, the harmonics of the induced dipole moment and 
the local electric field are both increased.
Similarly, in Fig.2 the same quantities are plotted but for the 
transverse electric field. We find that the harmonics are decreased as 
$\beta$ increases.
 
Both Fig.1 and Fig.2 also show that the harmonics are strongly 
dependent on the nonlinear response of the suspended particles.
Moreover, increasing the nonlinear response leads to an increase in 
the harmonics. 
This is consistent with the results of our recent work \cite{Wan},
in which a pair of nonlinear particles suspended in a linear host 
was considered.
  
In Figs.3 and 4, we compare the SC theory and PEM. 
For weak nonlinearity, $\chi E_0^2=0.09$, the PEM results
coincide with the SC results, while for moderate nonlinearity 
$\chi E_0^2=0.9$, there are some deviations.
In both cases, we find that the PEM results are in good agreement 
with the SC results for weak nonlinearity.
For strong nonlinearity, there are large deviations (not shown here) 
and the SC theory must be used.

The $\beta$ dependence of the local field can qualitatively be understood
with the spectral representation \cite{Bergman}. From Eq.(7), denoting
$s=(1-\ep_1/\ep_2)^{-1}$, we find that $E_1=sE_0/(s-s_1)$, 
where $s_1=(1-\beta f)/3$. Using the numerical values of
our model calculations, we find $s=-1/9$, and $E_1=(1/9)E_0/(1/9+s_1)$.
When $\beta$ increases, $s_1$ decreases, leading to an increase in the
local electric field $E_1$.

\section*{Discussion and conclusion}

Here a few comments on our results are in order. In the present study, 
we have examined the case of nonlinear particles suspending in a linear 
host. We may extend our considerations to a nonlinear host medium 
\cite{nler-2}. In this case, preliminary results show that qualitatively 
similar yet more complex behaviors in the ac response have been observed.

So far, we have not considered the frequency dependence of the particle
dielectric constant. In a realistic situation, the dielectric constant 
of the particles can decrease with the increase of the frequency. 
For simplicity, we may adopt the Debye relaxation expression for 
$\ep_1$. Preliminary results show that the ratio of the third to 
first harmonic decreases with frequency, results that are in accord with 
recent experimental data of Ref.\cite{Kling98}.

To our knowledge, an accurate evaluation of the local field factor 
$\beta$ during field-induced aggregation is lacking.
We suggest that the Ewald method \cite{Lo} can be extended to compute the
$\beta$ factor in computer simulation of ER fluids. 
This is a formidable task for the future.
Our present theory may be applied to the electrorheology of milk 
chocolate \cite{chocolate}. By measuring the nonlinear ac response, 
it may be possible to monitor the food production processes. 

In conclusion, a self-consistent formalism has been employed to
compute the induced dipole moment for suspensions in which the
suspended particles have a nonlinear characteristics, in an
attempt to investigate the anisotropy in the ac response.

\section*{Acknowledgments}
This work was supported by the Research Grants Council of the Hong Kong 
SAR Government under project number CUHK 4284/00P. K. W. Y. acknowledges
useful discussion with Prof. G. Q. Gu.

\begin{figure}[h]
\caption{The harmonic responses of the induced dipole moment and the 
 local electric field versus $\beta$ for the longitudinal field case:
 $\chi E_0^2=9$ (dotted lines), $\chi E_0^2=25$ (solid lines) and
 $\chi E_0^2=36$ (dashed lines).}
\end{figure}

\begin{figure}[h]
\caption{Same as Fig.1, but for the transverse field case.}
\end{figure}

\begin{figure}[h]
\caption{The comparison between the SC theory and PEM with 
 $\chi E_0^2=0.09$ and $\chi E_0^2=0.9$ for the longitudinal field case.}
\end{figure}

\begin{figure}[h]
\caption{Same as Fig.3, but for the transverse field case.}
\end{figure}

\newpage
\centerline{\epsfig{file=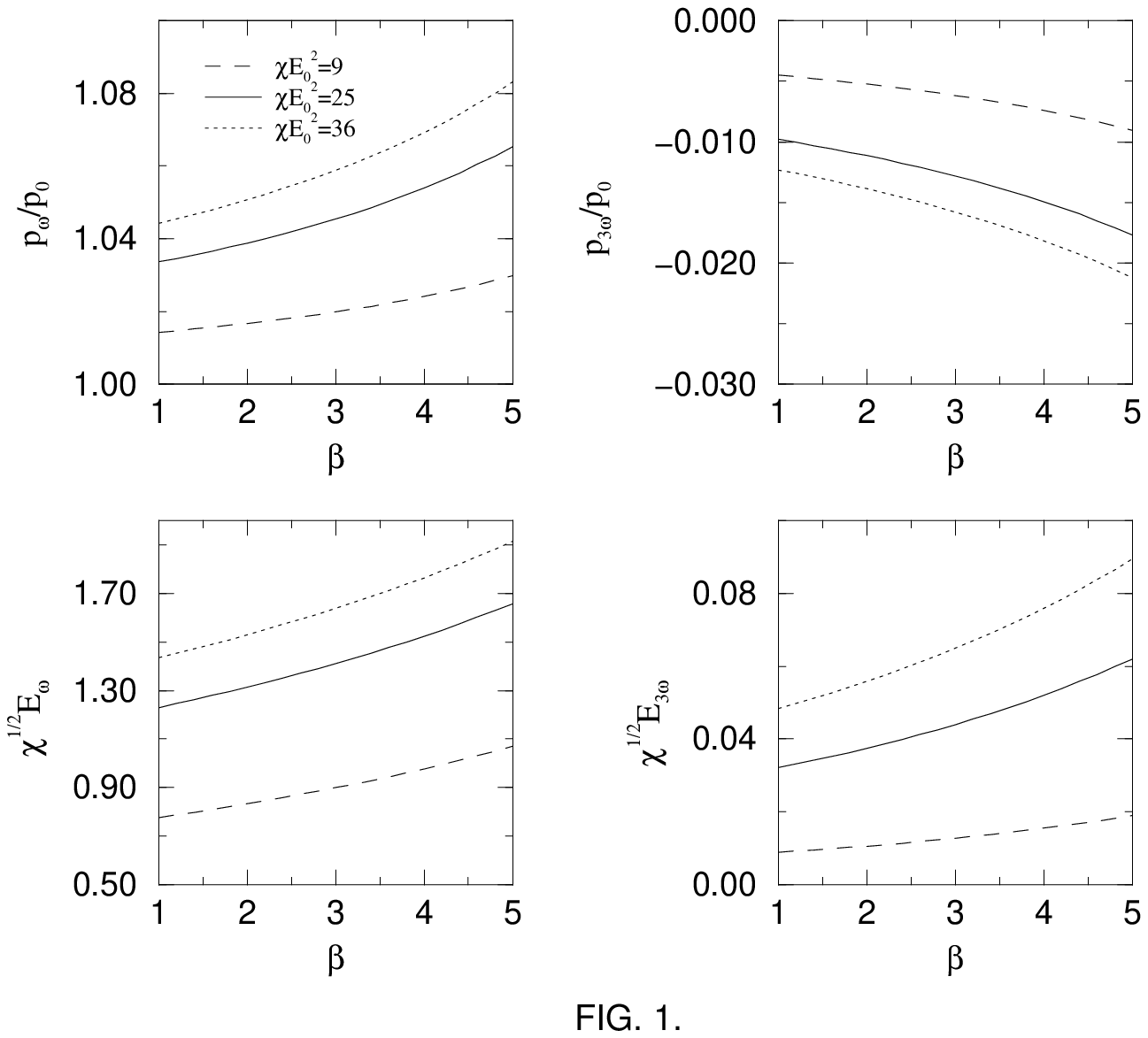,width=\linewidth}}
\centerline{Huang, Wan, Lo, Yu}

\newpage
\centerline{\epsfig{file=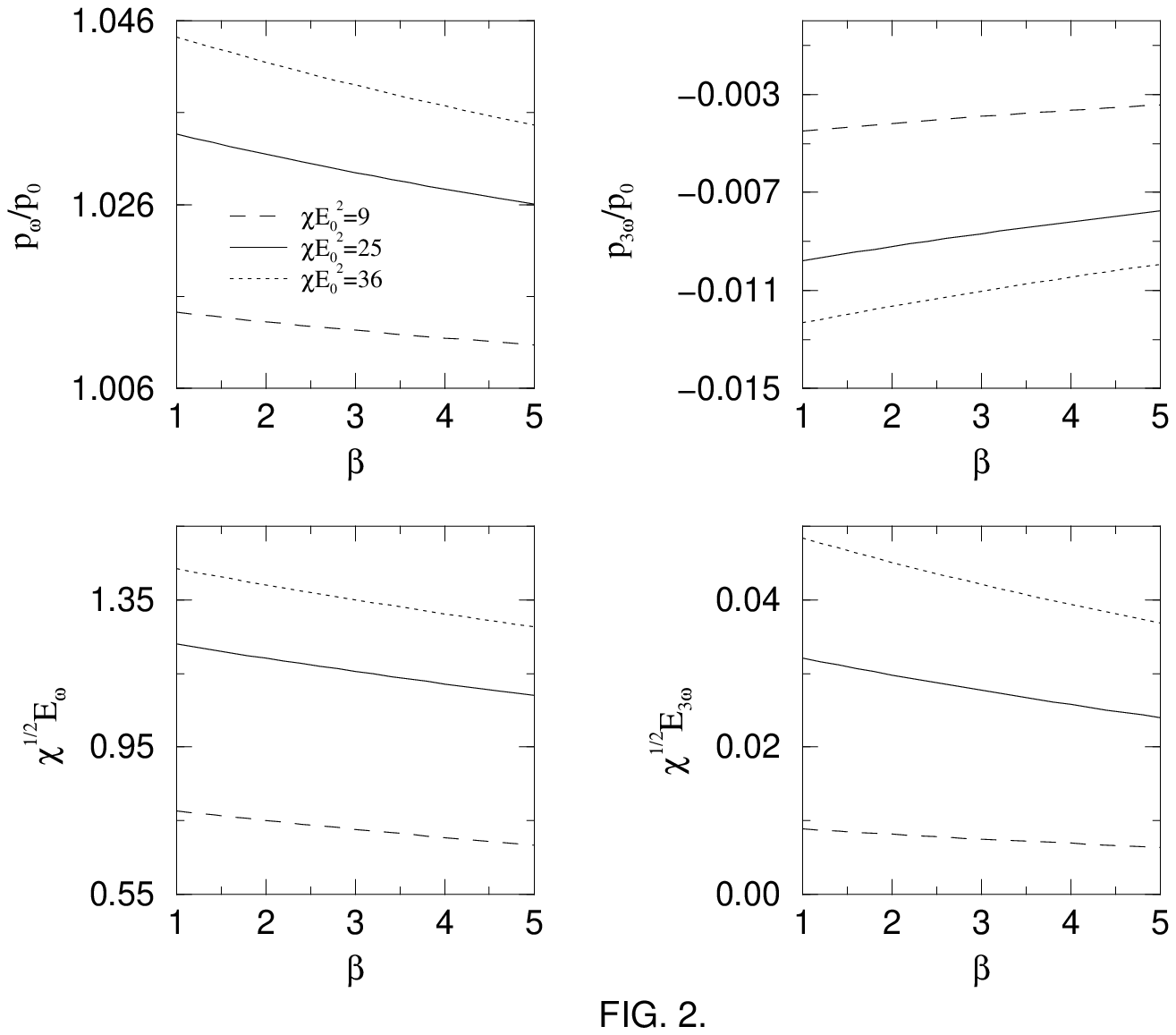,width=\linewidth}}
\centerline{Huang, Wan, Lo, Yu}

\newpage
\centerline{\epsfig{file=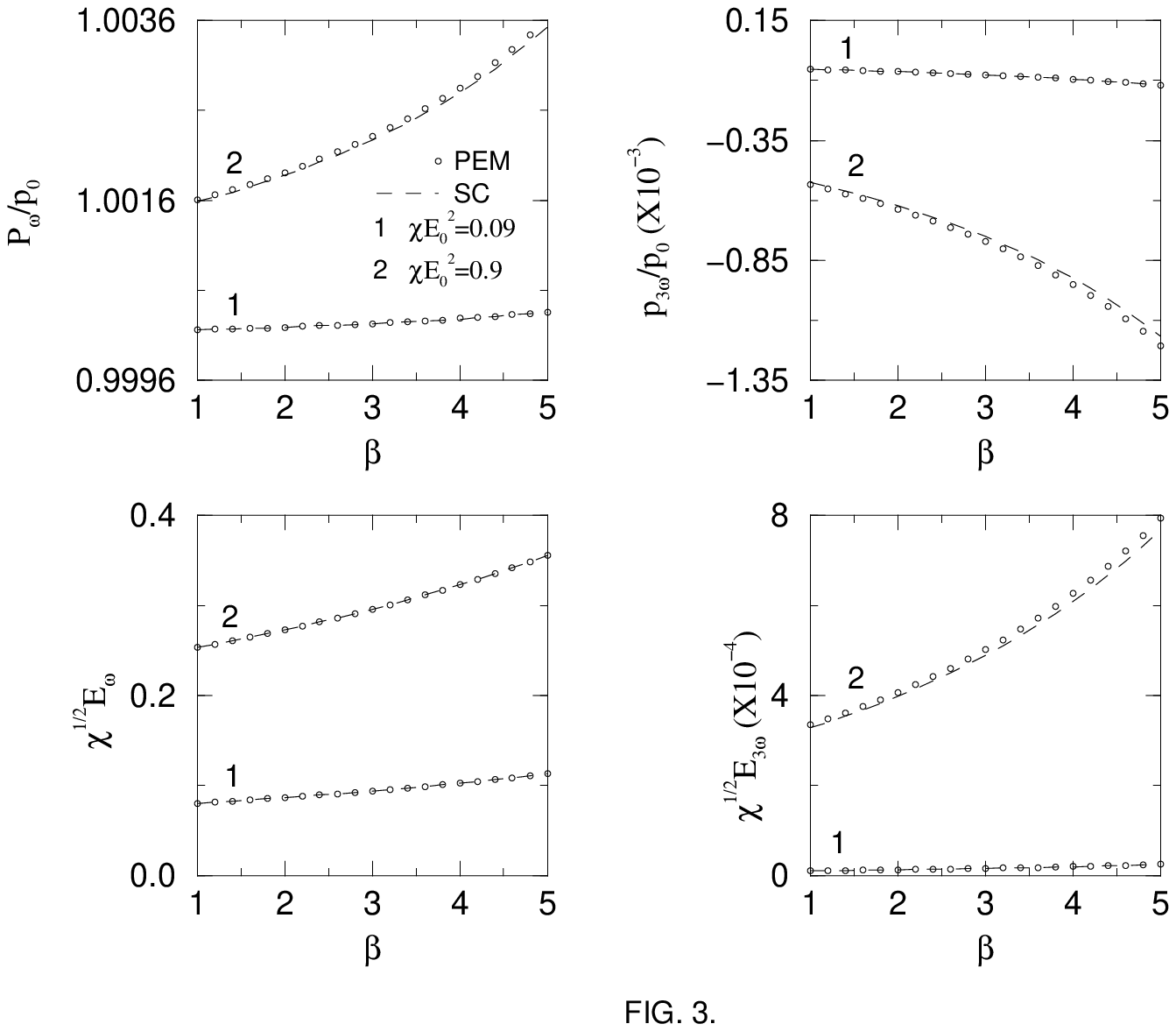,width=\linewidth}}
\centerline{Huang, Wan, Lo, Yu}

\newpage
\centerline{\epsfig{file=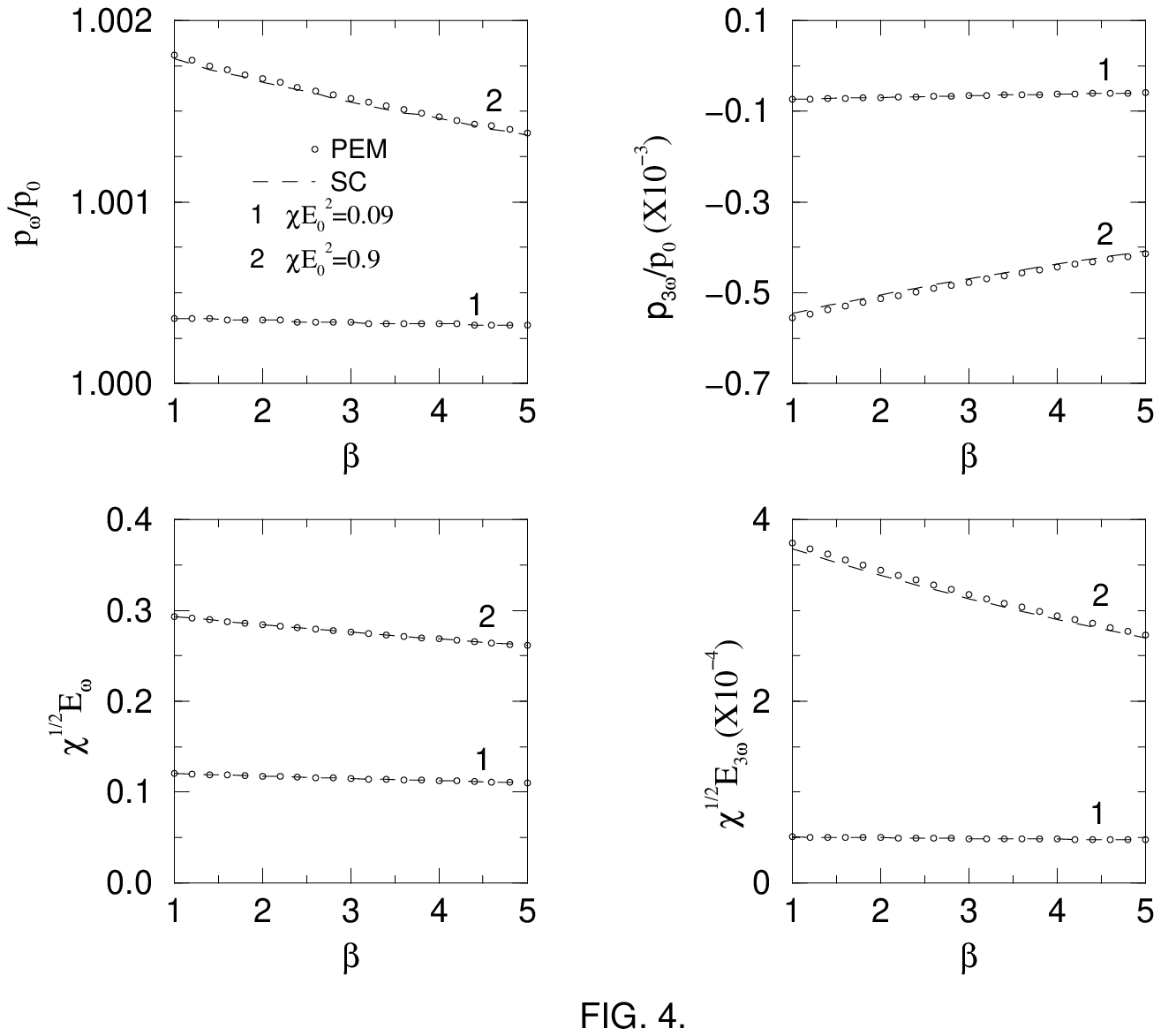,width=\linewidth}}
\centerline{Huang, Wan, Lo, Yu}

\end{document}